\documentclass[12pt,preprint]{aastex}

\usepackage{graphics,graphicx}
\usepackage{amsmath}
\usepackage{natbib}
\usepackage{color}
\def\vv{{\mathbf{v}}}
\def\vr{{\mathbf{r}}}
\def\vs{{\mathbf{J}}}
\def\vh{{\mathbf{h}}}

\def\phat{{\hat\psi}}

\newcommand{\gtorder}{\mathrel{\raise.3ex\hbox{$>$}\mkern-14mu
            \lower0.6ex\hbox{$\sim$}}}
\newcommand{\ltorder}{\mathrel{\raise.3ex\hbox{$<$}\mkern-14mu
            \lower0.6ex\hbox{$\sim$}}}

\shorttitle{Steady-State Transition Front}
\shortauthors{Hawley \& Krolik}

\begin{document}

\title{  A Steady-State Alignment Front in an Accretion Disk Subjected to Lense-Thirring Torques} 

\author{Julian H. Krolik\altaffilmark{1}}

\and

\author{John F. Hawley\altaffilmark{2}}

\altaffiltext{1}{Department of Physics and Astronomy, Johns Hopkins University, Baltimore, MD 21218, USA} 

\altaffiltext{2}{Department of Astronomy, University of Virginia, Charlottesville VA 22904, USA}

\begin{abstract}
Using only physical mechanisms, i.e., 3D MHD with no phenomenological viscosity, we have simulated
the dynamics of a moderately thin accretion disk subject to torques whose radial scaling mimics those produced by
lowest-order post-Newtonian gravitomagnetism.    In this simulation, we have shown how, in the presence
of MHD turbulence, a time-steady transition can be achieved between an inner disk region aligned with
the equatorial plane of the central mass's spin and an outer region orbiting in a different plane.
The position of the equilibrium orientation transition is determined by a balance between gravitomagnetic
torque and warp-induced inward mixing of misaligned angular momentum from the outer disk.
If the mixing is interpreted in terms of diffusive transport, the implied diffusion coefficient is
$\simeq (0.6$--$0.8)c_s^2/\Omega$ for sound speed $c_s$ and orbital frequency $\Omega$.
This calibration permits estimation of the orientation transition's equilibrium location given the central mass,
its spin parameter, and the disk's surface density and scaleheight profiles.   However, the alignment front overshoots
before settling into an equilibrium, signaling that a diffusive model does not fully represent the time-dependent
properties of alignment fronts under these conditions.   Because the precessional torque on the disk at the
alignment front is always comparable to the rate at which misaligned angular momentum is brought inward to the
front by warp-driven radial motions, no break forms between the inner and outer portions of the disk in our simulation.
Our results also raise questions about the applicability to MHD warped disks of the traditional distinction between
``bending wave" and ``diffusive" regimes.
\end{abstract}


\section{Introduction}

It is now forty years since \citet{BP75} pointed out that the precessional torques driven by the Lense-Thirring
effect would, when applied to a fluid disk surrounding a spinning black hole, lead to alignment of its inner
parts' orbital axis with the black hole's spin, while leaving untouched its original orientation, perhaps sharply misaligned,
at large radius.   Despite this long passage of time, there remains controversy over such basic issues as the detailed
mechanism of alignment, where the transition between orientations occurs, and how its location depends
on such disk properties as the disk's sound speed or surface density distribution.    \citet{BP75} suggested
that the orientation transition would be determined by a competition between the Lense-Thirring torques
and the importation of fresh misaligned angular momentum by the accretion flow.   \citet{PP83}
recognized that misaligned angular momentum can be more rapidly transported by the radial flows
induced by disk-warping than by accretion, but the nature and speed of these radial flows was left uncertain
because their regulation was ascribed to an imagined isotropic viscosity, which was assumed to be related to
the local pressure through a universal dimensionless
constant, $\alpha$.   \cite{Pringle92} developed a simpler formalism to describe radial flow of misaligned angular momentum,
in which it was supposed to spread diffusively, but the ``diffusion coefficient" relating a radial gradient
of angular momentum orientation to an associated angular momentum flux was estimated assuming that
radial flows are controlled by that same imagined viscosity.    \citet{O99} introduced a more complicated version
of this approach by creating a quasi-linear theory of this model, one in which the angular momentum orientation diffusion
coefficient became a function of several parameters.   \citet{NP00} constructed a formalism for
relating the rate of radial transport of misaligned angular momentum to the location of the orientation
transition, but in the absence of a clear understanding about how the radial flows were regulated,
this formalism could not yield a definite statement about its location.   Recently, \cite{Nixon2012a}, also working
within the isotropic viscosity formulation, pointed out that large misalignment
angles might lead to an actual rupture of the disk, rather than a smooth transition.   Shortly after, detailed
hydrodynamics simulations \citep{SKH13a} showed that the relation between angular momentum flux and disk
warp has clear departures from the diffusion model.    

This situation can now be clarified.   There is a well-established physical mechanism for internal stresses
in accretion disks---correlated magnetohydrodynamic (MHD) turbulence, driven by the magneto-rotational
instability \citep{mri91,bh98}.
In addition, although treating MHD turbulence in disks requires very demanding numerical simulations,
computer algorithms and computer power are now at a level such that even warped precessing disks can be treated directly.
Disks tilted with respect to the rotation axis of a central black hole have been simulated in fully relativistic three
dimensional MHD \citep{Fragile07,McK2013,Fragile14a}, but the computational expense of full general relativity forced
parameter choices (disk thickness, radial extent) that made accretion-related alignment processes difficult to see.
The central difficulty is that if a numerical simulation is to follow the MHD turbulence, it must  have a timestep
very short compared to an orbital timescale, whereas the precession timescale where the orientation transition
may occur is almost certainly many orbital periods long.   Consequently, simulations able to probe the orientation
transition at a realistic scale are prohibitively costly.

However, it is possible to avoid this roadblock by
using semi-Newtonian methods, in which the only relativistic effect retained is the lowest-order post-Newtonian
term describing Lense-Thirring torque.     Progress is possible in this framework because Newtonian gravity is
both scale-free and obeys a linear field equation.  As a result, in this treatment there is no maximum
value for the spin parameter.    Simulators are then free to choose the precession frequency associated with
the Lense-Thirring torques, constrained only by the provisos that it must have the proper radial dependence,
and it must be slower than the orbital frequency.

By adopting this device, \citet{SKH13b} were able to simulate a disk in which an alignment front propagated
steadily outward.   By studying this model in detail, they were able to achieve several things.   First,
they demonstrated that the internal MHD stresses triggered by disk warp are fundamentally not viscous in character
and have a magnitude much smaller than predicted by the ``isotropic $\alpha$" model (\cite{Fragile14a} reached
a similar conclusion).    Second, although the magnetic forces associated with these stresses are generally weaker than
the fluid pressure forces, the turbulence created by MHD effects significantly influences disk dynamics because
it disrupts the propagation of bending waves.   Third, they identified the specific mechanism by which alignment occurs:
when the precession angle of the disk decreases outward (as it in general should for Lense-Thirring torques), the torque
has a component that is in part opposed to the misaligned angular momentum at larger radius than where it is
exerted.   If the new angular momentum delivered to the disk is then transported outward by radial fluid motion,
it progressively cancels the misalignment.   Fourth, based on this understanding of the alignment mechanism,
they developed a simple model for the speed of propagation of the alignment front that fit their data very well.   However,
because the radial surface density profile chosen for the simulated disk had
a peak in the middle of the disk, there was too little misaligned angular momentum stored in its outer regions
to be able to slow, much less stop, the outward propagation of the alignment front.   Thus, \citet{SKH13b} were unable
to study a time-steady orientation transition.

In this paper, we present a new simulation, very similar in most respects to that of \citet{SKH13b}, but with
a surface density distribution altered so that the majority of the disk's mass is located at large radius.
This choice creates a situation in which the alignment front can travel outward, but then
decelerate and stop.   This is just what occurs, and in this paper we report in detail on the properties
of such a time-steady orientation transition.

\section{Simulation Details}
\label{sec:sims}

As in \citet{SKH13b}, we use our Fortran-95 version of the 3D
finite-difference MHD code {\it Zeus} \citep{zeus1,zeus2}.  The {\it Zeus}
code solves the standard equations of Newtonian magnetohydrodynamics
using direct finite differencing.  The magnetic field is updated using
the method of characteristics constrained transport (MOCCT) algorithm
to maintain zero divergence to machine accuracy \citep{HawleyStone95}.

Our procedure is to begin with a conventional flat disk, evolve it long enough to allow the MRI-driven
MHD turbulence to saturate, and only then turn on a torque-driving Lense-Thirring
precession.   To represent this torque, we add a force term of the form $\rho \vv \times \vh$
to represent the gravitomagnetic force per unit mass, where $\rho$
is the mass density, $\vv$ is the fluid velocity, and
\begin{equation}
\vh = \frac{2\vs}{r^3} - \frac{6(\vs \cdot \vr)\vr}{r^5}.
\end{equation}
Here $\vs$ represents the magnitude and direction of the spin vector of the central mass
and $r$ is spherical radius.   We choose the direction of the spin vector to be in the
$x$-$z$ plane, tilted $12^\circ$ (0.21 radians) from the $z$-axis in the $\hat x$ direction.
Its magnitude is such as to cause a test-particle at $r=10$, our fiducial radius, to precess at a
frequency $1/15$ of the orbital frequency at that radius.

The simulation grid uses spherical coordinates with a spatial domain
in $(r,\theta,\phi)$ that spans $ [4,40] $ in radius,  $[0.1,0.9]\pi$
in $\theta$ and a full $ 2\pi $ in $\phi$.  Since we are using
Newtonian gravity rather than (say) a pseudo-Newtonian gravity with a special lengthscale
corresponding to the gravitational radius $r_g = GM/c^2$, the units of length are arbitrary.
For consistency with \citet{SKH13b} we use the same units, in particular
measuring time in units of the orbital period at $r=10$, and setting $GM=1$.
We use  $(352,384,1024)$ grid cells in the radial, polar, and azimuthal
directions respectively. The radial mesh is spaced  logarithmically with
a constant $\Delta r/r = 0.00654$. We employ a polynomial spacing in the polar
dimension (Eqn.~6 of \citet{NKH10}, with $\xi = 0.65$ and $n = 13$).
This polynomial spacing focuses cells near the equatorial plane,
placing 64\% of the $\theta$ zones within $\pm 10^\circ$ of
the midplane.  The $\phi$ zones are uniform in size.  Outflow boundary conditions are employed
on the radial inner and outer boundary, and along the $\theta$ boundary that forms a ``cut-out'' around the polar axis.

In our initial condition, the fluid orbits a point-mass in Newtonian gravity with a
Keplerian angular velocity, $\Omega^2 = GM/r^3$.   A similar initial disk was used
in \cite{Jiming}.   Its equation of state is isothermal, with sound-speed $c_s^2 =
0.001$, which corresponds to 0.1 of the orbital velocity at the fiducial
radius of $r=10$.   The scale height of the disk, $H=c_s/\Omega$, is then $(r/10)^{3/2}$.
If the scale height is instead defined in terms of a vertical density moment, the disk aspect
ratio $h/r \simeq 0.07$ at $r \simeq 6$ and rises to $\simeq 0.12$ at $r \simeq 20$.   The disk's
opening angle is thus about half the initial tilt of the disk relative to the black hole spin.
We set the density $\rho_c = 1$ at the equator at all radii, and the
vertical distribution of density is $\rho = \rho_c \exp (-z^2/2H^2)$.   The surface density
$\Sigma$ increases with radius from the disk's inner radius $r_{\rm in} = 6$ out to
$r = 20$ and declines outward from there to its outer radius $r_{\rm out} = 28$
(see Fig.~\ref{fig:surfdens_early} for $\Sigma(r)$ when the torque is turned on).
At the inner and outer limits the disk is simply truncated. The
initial condition is not in radial pressure equilibrium (especially at the disk
boundaries).  

The initial magnetic field in the disk is a large set of nested dipole loops extending
from the inner to the outer portions of the disk.  The field is defined by a vector potential
\begin{equation}
A_{\phi} = A_0 \rho^{1/2} \sin (2\pi r/\lambda H) (r/r_{in} -1) (1-r/r_{out}) .
\end{equation}  
The scalefactor $\lambda$ is set equal to $2/c_s$.
The vector potential is limited to positive values with a cutoff at $0.05\rho_c$, i.e.,
\begin{equation}
A_{\phi} = \max(A_{\phi} - 0.05 \rho_{c},0).
\end{equation}  
The field amplitude factor $A_0$ is chosen so that the volume-integrated ratio of gas
to magnetic pressure, the plasma $\beta$,  is 1000.
Turbulence is seeded by imposing random pressure perturbations at the 1\% level on the initial condition.

Throughout the simulation we monitor a number of quantities in order to gauge the numerical 
quality of the simulation.  These metrics were developed and studied in 
\cite{HGK11,HRGK13}, as well as in \cite{Sorathia12}. 
One set of metrics measures the ratio of the characteristic MRI wavelength $\lambda_{MRI}$ 
to the grid resolution.  Specifically,
\begin{equation}\label{eq:qtheta}
Q_\theta = {\lambda_{MRI} \over r \Delta \theta},
\end{equation}
and
\begin{equation}\label{eq:qphi}
 Q_\phi = {\lambda_{MRI}\over r \sin\theta \, \Delta\phi }.
\end{equation}
The characteristic MRI wavelength is defined as $\lambda_{MRI} = 2\pi |v_A|/\Omega$, 
where the Alfv\'en speed $v_A$ is 
obtained from the appropriate component of the magnetic field ($B_\theta$ for $Q_\theta$, $B_\phi$
for $Q_\phi$).  The $Q$ numbers are then
the number of grid cells that span the characteristic wavelength of the MRI.  \citet{HGK11} and \citet{HRGK13}
estimate that $Q$ values $>15$--$20$ are indicative of adequate resolution. For the initial
magnetic field given here, $\lambda_{MRI} \sim 0.1$, and the initial $Q_\theta$ values at the
disk equator inside the fiducial radius range from $\sim 4$--8.  There is no initial toroidal 
field, but orbital shear creates toroidal field from radial field on the orbital timescale.   Thus,
within a single orbit we can expect $B_\phi \sim B_r$, producing an azimuthal quality factor
$Q_\phi = 2$--15 at $r=10$.   Although this would seem like marginal resolution, the $Q$ values
increase as $B$ grows beyond the initial field amplitude (see below).

A second set of metrics measures the average properties of fully developed MHD turbulence;
these are, by definition, irrelevant in the initial condition of the simulation, so we do not evaluate
them until the turbulence develops.   These metrics
are calibrated by the values measured in highly-resolved local simulations.  
\cite{HGK11,HRGK13} develop two such diagnostics, $\alpha_{mag} = M_{r\phi}/P_{mag}$, 
the ratio of the Maxwell stress $M_{r\Phi}$  to the magnetic pressure, and 
$\langle B_r^2\rangle/\langle B_\phi^2 \rangle$,
the ratio of the radial to toroidal magnetic energy.
When suitably averaged over the computational domain in well-resolved simulations, these quantities 
approach limiting values of $0.45$ and $0.2$, respectively \citep{HRGK13}.

\section{Results}
\label{sec:results}

\subsection{Initial disk evolution}

The aim of our new simulation is to follow the process by which an accretion disk subjected
to Lense-Thirring torques reaches a steady-state in which its inner portions are aligned with
the central black hole while its outer portions remain oblique.   Because the character of this
process depends upon the fluctuations created by MHD turbulence \citep{SKH13b}, we begin by
evolving the initial disk until it is turbulent throughout most of its radial extent.  Only after that state is
achieved do we switch on the torque.

Because the initial disk was not in radial pressure equilibrium, a period of adjustment occurs as
the disk evolves.  Although the unbalanced  pressure gradients are small within the body of the disk
due to the small ratio of sound speed to orbital speed, they are large at the disk boundaries.  Initially,
the inner edge of the disk pushes inward, but it then rebounds, driving a strong compression
wave through the disk, while the outer radius of the disk expands outward,
sending a rarefaction wave into the inner disk.

By $t\simeq 12$ fiducial orbits, which marks the end of the torque-free evolution, turbulence has been well
established in most of the disk.   Inside $r=20$ at this time, the mass-weighted mean value of $Q_\theta$
varies between 15 and 22, $Q_\phi$ is between 35 and 50,
$\alpha_{mag}$ has an average value of 0.37, and $\langle B_r^2\rangle/\langle B_\phi^2\rangle$
is 0.16.  The two $Q$ values are well above the resolution thresholds set by \cite{HGK11,HRGK13}, while the
$\alpha_{\rm mag}$ and  $\langle B_r^2\rangle/\langle B_\phi^2\rangle$ values are a little bit lower than
the limiting values found in that paper.

\begin{figure}
\begin{center}
\includegraphics[width=0.6\textwidth,angle=90]{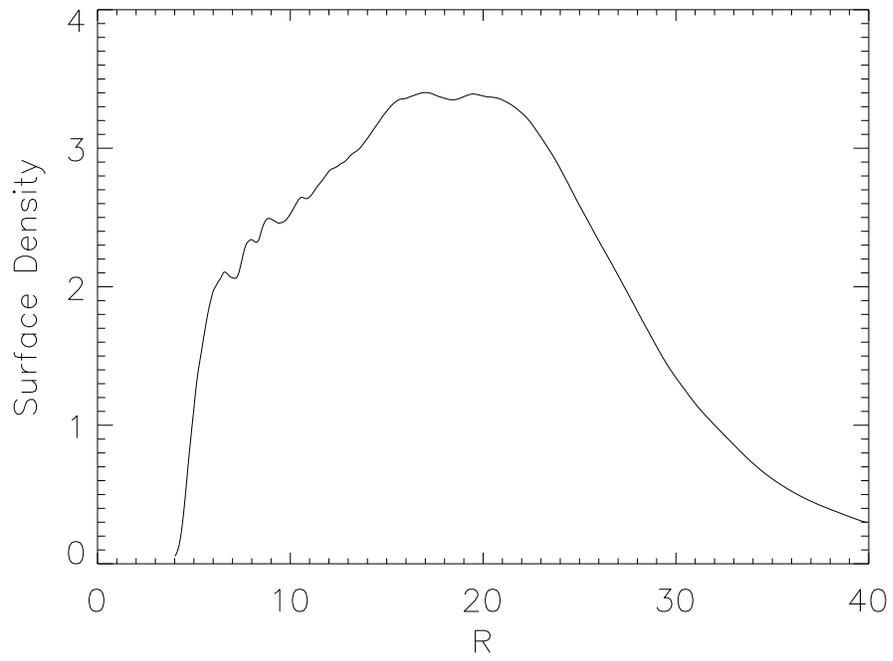} \\
\caption{Surface density as a function of radius at $t=12.4$, the beginning of the torque.}
\label{fig:surfdens_early}
\end{center}
\end{figure}

Under the influence of the magnetic stresses associated with the turbulence, the disk spreads radially,
both inward and outward.   At the end of the torque-free evolution period, the disk has 65\% of its original
mass, and a surface density with the form shown
in Figure~\ref{fig:surfdens_early}.   Unlike the radial surface density profile of the disk studied by \citet{SKH13b},
which reached a peak at $r \simeq 12$ and declined sharply from there outward, the surface density in
this simulation rises all the way to $r \simeq 20$, reaching a peak roughly twice the surface density
prevailing for $6 \lesssim r \lesssim 10$.     Thus, more than enough ``misaligned" angular momentum
$\vec L_{\perp} \equiv \vec L - \vec L \cdot \hat s$, with $\hat s$ the direction of the nominal black hole
spin, is present on the grid to hold back the outward progress of an alignment front for a long time.

\subsection{Evolution with torque}\label{sec:torque_evol}

At 12.4~fiducial orbits, we turned on the torque and then ran the simulation for another 12.4~fiducial
orbits.   Because the black hole spin implied by the torque is tilted $12^\circ$ from the polar axis of
the coordinates, the disk acquires an effective misalignment angle $\beta = 12^\circ$ at this point
in the simulation; at subsequent times, we define the radius-dependent misalignment angle
$\beta(r) \equiv \arcsin \left[L_\perp(r) /L_{\rm tot}(r)\right]$, where $L_{\rm tot}(r)$ is the magnitude of the
shell-integrated angular momentum at radius $r$, and $L_\perp(r)$ is the magnitude of its component
in the plane perpendicular to the black hole spin.

\begin{figure}
\begin{center}
\includegraphics[width=0.6\textwidth,angle=90]{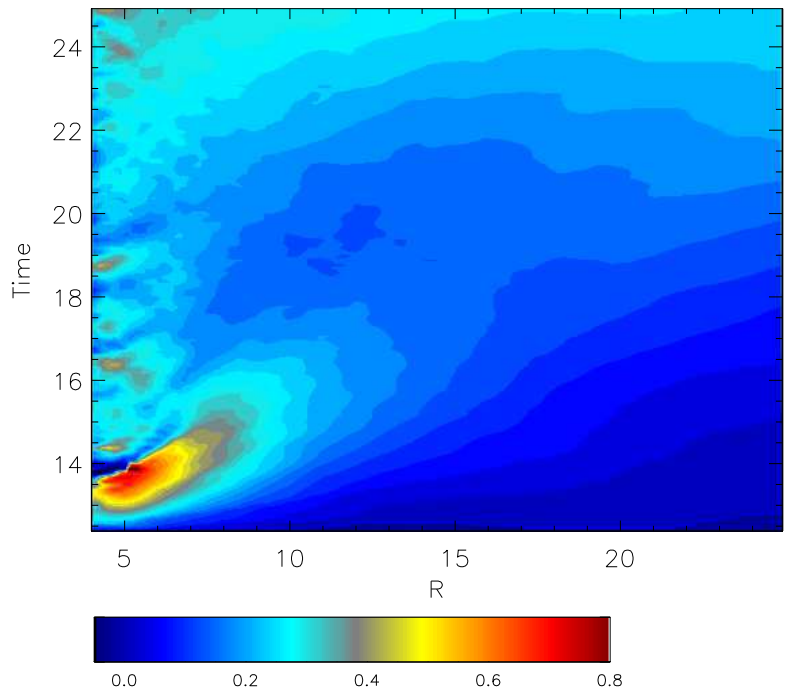} \\
\caption{Color contours (see color bar) of the precession angle $\phi_{\rm prec}/\pi$ as a function of radius and time.
Although the simulation extended to $r=40$, we show only the range $4 \leq r \leq 25$ because little happens
at larger radius.   The precession angle in the small region at $r \lesssim 7$ and $t \simeq 14$ is ill-defined
because the disk is almost exactly aligned (see Fig.~\ref{fig:align}).}
\label{fig:precess}
\end{center}
\end{figure}

If orbiting material moved as independent test-particles, the inner rings would precess much faster than the outer
rings because the precession frequency declines outward $\propto r^{-3}$.   To measure the actual precession,
we define the precession angle $\phi_{\rm prec} \equiv \tan^{-1} (L_{y}^\prime/L_{x}^\prime)$, where $L_{x,y}^\prime$
are the shell-integrated $x$ and $y$ components of the angular momentum relative to a coordinate system found
by rotating the grid coordinates around the $y$ axis in order to make the black hole spin define the $z$ axis.
As Figure~\ref{fig:precess} shows, although the inner rings do precess faster, the radial gradient is noticeably
shallower than $r^{-3}$: after 1~orbit, although a test-particle at $r=5$ would have precessed by $\simeq 1.1\pi$,
the actual fluid ring at that radius has precessed by only $ \simeq 0.6\pi$, while the ring at $r=10$ has
precessed by an amount similar to the test-particle prediction; after 3~orbits, precession is ill-defined at $r=5$
because that portion of the disk has already almost completely aligned, but the difference in precession angle
between $r=10$ and $r=15$ is only $\simeq 0.1\pi$, in contrast to the test-particle expectation of $\simeq 0.3\pi$.

Several other features in this diagram are even more contrary to test-particle expectations.   Most importantly, the
precession angle hardly changes from $t \gtrsim 15$ onward for $r \lesssim 10$.   Also beginning at
$t \simeq 15$, or after only $\simeq 2.5$ orbits of torque, a zone at larger radius, $7 \lesssim r \lesssim 13$,
{\it reverses} its sense of precession for several orbits.   Once the sense of precession at large radius resumes
its normal sense (at $t \simeq 20$), the precession phase gradient becomes very shallow throughout the disk.

Meanwhile, the innermost disk aligns very swiftly, within a few fiducial orbits.   The location of the alignment front
moves rapidly outward, with the half-alignment point reaching $r \simeq 10$ only $\simeq 3.5$ fiducial orbits
after the torques begin.   However, there is a turn-around in the process and the front retreats over the next several
orbits until it reaches its steady-state position at $r \simeq 7$--8.    From $t \simeq 20$ until the end of the simulation
at $t \simeq 25$, neither the position nor the structure of the front changes appreciably.

\begin{figure}
\begin{center}
\includegraphics[width=0.6\textwidth,angle=90]{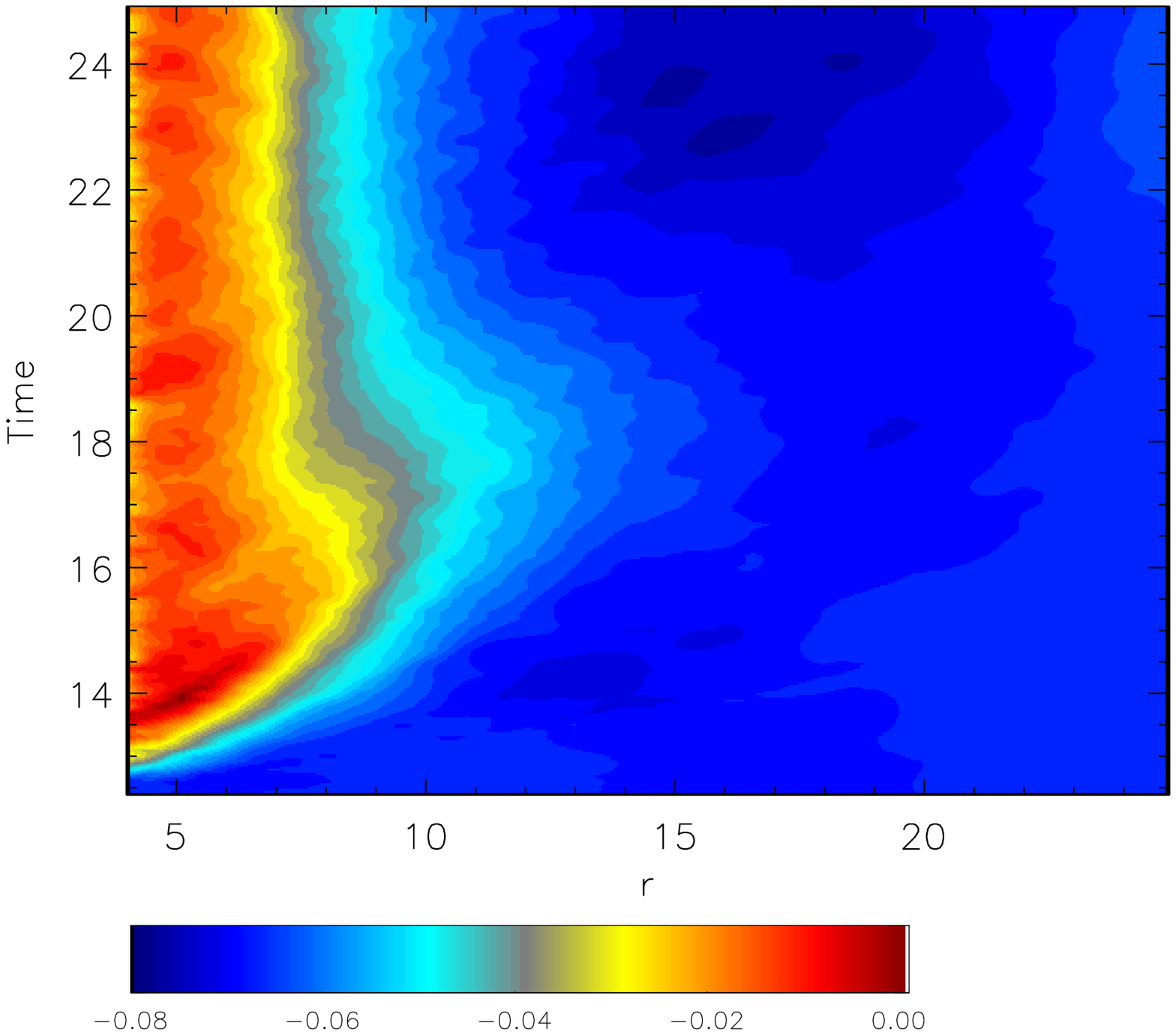} \\
\caption{Color contours (see color bar) of the misalignment angle $\beta/\pi$ as a function of radius and time.
Although the simulation extended to $r=40$, we show only the range $4 \leq r \leq 25$ because little happens
at larger radius.}
\label{fig:align}
\end{center}
\end{figure}

\begin{figure}
\begin{center}
\includegraphics[width=0.6\textwidth,angle=90]{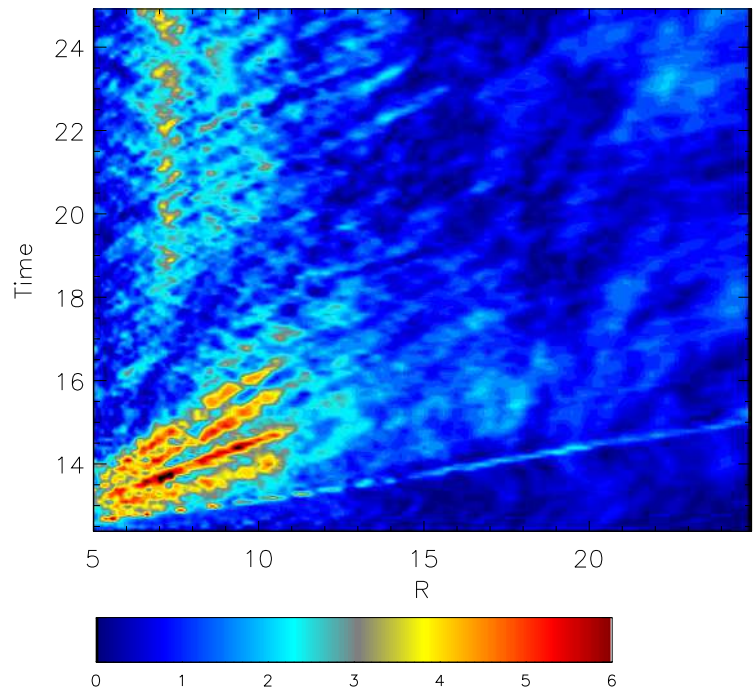} \\
\caption{Color contours (see color bar) of the normalized disk warp $\hat\psi$.}
\label{fig:psihat}
\end{center}
\end{figure}

Even though the disk remains near its original alignment at $r \gtrsim 12$ after $t \simeq 20$, it
continues to precess slowly at large radii, turning at a rate $\simeq 0.03\pi$ radians per fiducial orbit, roughly
the test-particle precession rate at $r \simeq 16$.    Although there is little torque inside the alignment radius ($r \simeq 7$),
there is sufficient obliquity at larger radii that the disk continues to feel a torque.   This torque is spread to larger
radii by the same mechanisms seen in earlier studies \citep{SKH13a,SKH13b}.   When the radial precession
gradient creates an inter-ring warp of nonlinear amplitude, it triggers a warp pulse that travels outward.   Nonlinear
in this context means that the parameter $\phat \equiv |d\hat \ell/d\ln r|/(h/r) > 1$, where $\hat \ell (r)$
is the unit vector in the direction of the total angular momentum associated with the shell at radius $r$.   When
that criterion is satisfied, the warp creates an order-unity radial pressure contrast.    The resulting transonic
fluid motions mix angular momentum between rings, smoothing the warp.   This same smoothing process
keeps the pulses narrow.  Two groups of such pulses can be clearly seen in Figure~\ref{fig:psihat}.  During the first
$\sim 3$~orbits, five pulses of strong ($\hat\psi \sim 4$--6) warp propagate outward.   
Although each one travels at very nearly constant speed, the pulse speed declines from first to last, falling from
$\simeq 8$ to $\simeq 2$, i.e., from $\simeq 1.3c_s$ to $\simeq 0.3c_s$, from first pulse to last.   Their direct
impact on local precession angle can be seen immediately by comparing Figures~\ref{fig:precess} and \ref{fig:psihat}.
A second group, somewhat weaker ($\hat\psi \sim2$--3), are visible in the zone $8 \lesssim r \lesssim 15$ and
$20 \lesssim t \lesssim 25$.   Although less dramatic, their effect on precession phase can likewise be seen
by examination of the region in radius--time space occupied by these pulses in Figure~\ref{fig:precess}.
All of this group of pulses have rather similar speeds, $\sim 3$~length units per fiducial orbital period, $\simeq 0.5c_s$.
As hinted by this speed, these pulses are related to linear bending waves, whose characteristic propagation speed is half the
isothermal sound speed \citep{Lubow02}.   However, a number of factors combine to make their
behavior somewhat different.     In addition to the rapid local damping caused by their nonlinearity, Figure~\ref{fig:psihat}
also shows that their amplitude is quite irregular, a symptom of their propagation through a background medium
made turbulent by the magneto-rotational instability \citep{SKH13b}.

\begin{figure}
\begin{center}
\includegraphics[width=0.6\textwidth,angle=90]{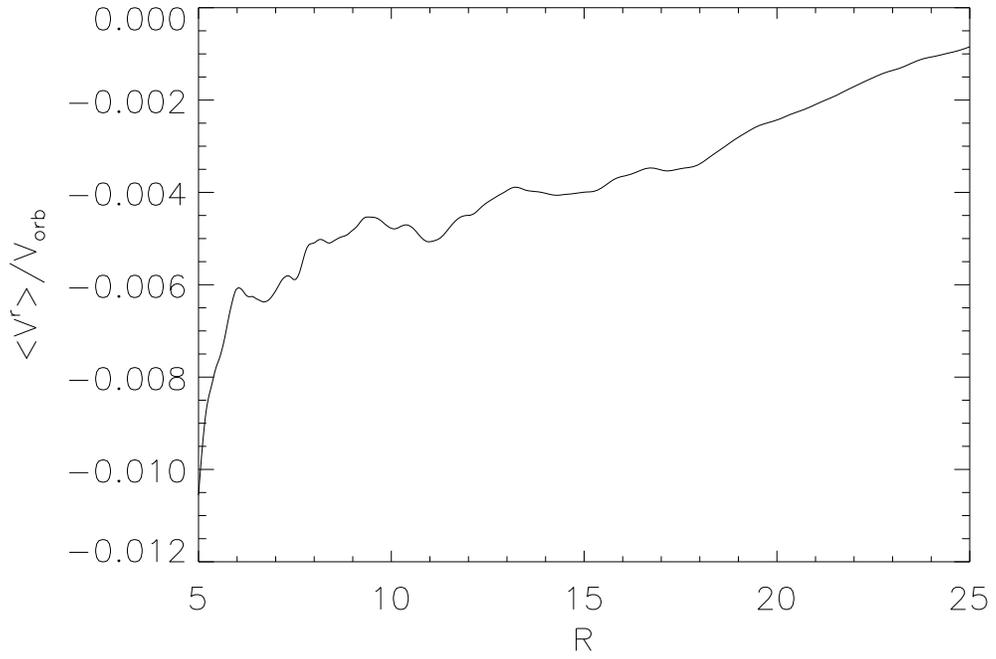} \\
\caption{Ratio of radial velocity to orbital speed as a function of radius.   Negative radial velocity indicates inward motion.}
\label{fig:vrad}
\end{center}
\end{figure}

\begin{figure}
\begin{center}
\includegraphics[width=0.6\textwidth,angle=90]{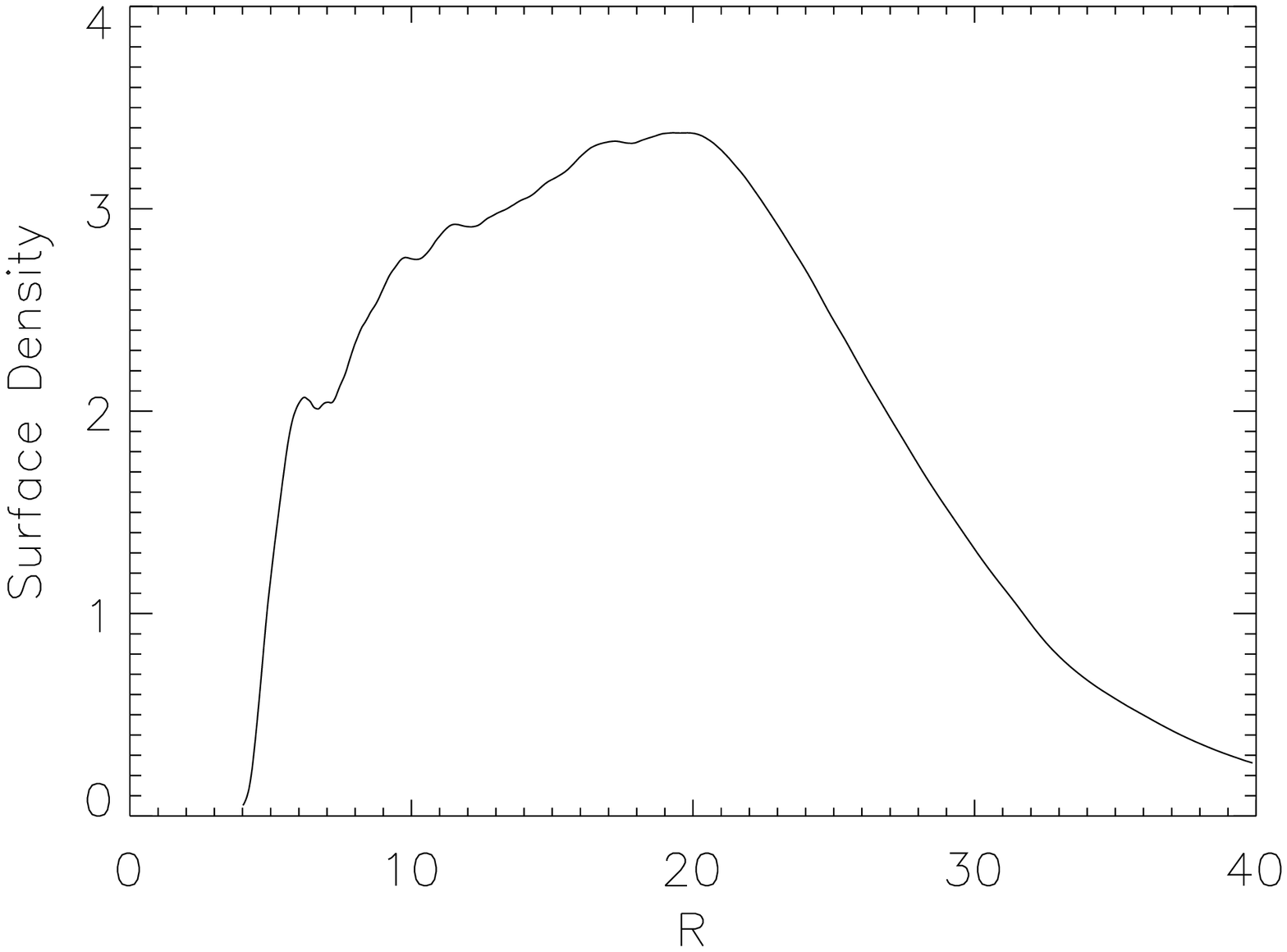} \\
\caption{Surface density as a function of radius at $t=28$, shortly before the end of the
simulation.}
\label{fig:surfdens_late}
\end{center}
\end{figure}

Throughout the evolution, the disk continues to accrete due to internal stresses, but in the vicinity of the
alignment front, the radial motions induced by the disk warp create a large enough
Reynolds stress to enhance the local inflow rate.  To measure this effect,  we compute the
mean radial, i.e., accretion, velocity from the mass flux as follows:
\begin{equation}
\langle v^r \rangle = \frac{\int \rho v^r r^2 \sin\theta d\theta d\phi}{ \int \rho r^2 \sin\theta d\theta d\phi} .
\end{equation}
This velocity is then averaged over time for the last orbit of the simulation. The result is
shown in Figure~\ref{fig:vrad} as the ratio of the accretion velocity to local orbital velocity.
This ratio is $\simeq -0.004$ well outside the transition front  ($13 \lesssim r \lesssim
18$), but its magnitude increases inward from that radius, becoming $\sim 50\%$ greater
by $r \simeq 7$, the location of the front.  Another consequence of this effect is a shift in
mass from the surface density peak, which is at $r \simeq 15$--20 when the torque begins,
to the transition front region (compare Figs.~\ref{fig:surfdens_late} and \ref{fig:surfdens_early}).
Note that the diminution in radial speed at very large radius is an artifact common to
any accretion simulation with a finite disk: matter on the outside must move still farther away as
it absorbs angular momentum transported to larger radius by the accretion stresses.

Over the course of the evolution with torque, the quality parameters $Q_{\theta,\phi}$ maintain values consistent
with adequate resolution.  Within the region of the disk running from $r=6$--20, $Q_\theta$ ranges from
9--14, while $Q_\phi$ varies between 20 and 30.

\section{Analysis}

\subsection{Overshoot}

Because the torque is turned on abruptly, the disk is subject to some transient effects as it adjusts;
one of these is an overshoot in the location of the alignment radius.  
This ``overshoot"---the quick progress outward of the alignment front followed by a retreat to its
long-term position---can be understood as a combination of several previously-identified processes.

The overshoot begins as the five strong warp pulses described in the previous section propagate outward.
When the inner regions are advanced in precession phase relative to the outer regions, the gravitomagnetic
torque delivered to the inner regions is directed so that it can partially cancel the misaligned angular momentum
of the outer regions \citep{SKH13b}.    Accomplishing this cancellation requires transport outward, and the pulses
do this.   Alignment by this mechanism can be particularly rapid when there is little warp at still larger radii because
radial mixing flows bringing inward angular momentum aligned with the original orientation of the disk are weak.

However, a consequence of quick alignment is the creation of significant warp.   After roughly a local
orbital time (and at the relevant radii near $r \simeq 10$, the local orbital time is the same as our
fiducial orbital period), the newly-created warp accelerates radial mixing flows \citep{SKH13a}.   The delay
time is roughly a local orbital period because this is the local dynamical time.  These radial flows carry
inward angular momentum that is both more misaligned and further behind in precession phase.
As a result, the local alignment is reduced and the precession temporarily reverses.   Thus, both
the alignment front overshoot and the precession reversal can be understood as consequences
of transients associated with the beginning of torque in a flat, but misaligned, disk.  Although it is somewhat
artificial to turn on the Lense-Thirring torque abruptly, this procedure does allow us to highlight the 
dynamic alignment process as it subsequently occurs.  The disk ultimately achieves a steady
state alignment configuration, despite the impulsive initial nature of the torque.

\subsection{Location of the steady-state orientation transition}

The ultimate location of the alignment front is also very consistent with the formalism proposed by
\citet{SKH13b} to estimate both the propagation speed of an alignment front and the location where
it stalls, reaching a steady-state.   They proposed that the alignment front should move outward
at a rate
\begin{equation}
\frac{d r_f}{dt} = \frac{2 \langle \cos\gamma \rangle a_* (GM)^2}{\sin\beta(r_f) c^3 r_f^{3/2} \Sigma(r_f)}
 \int_0^{r_f} \, dr^\prime \sin\beta(r^\prime)\Sigma(r^\prime)/{r^\prime}^{3/2} ,
\end{equation}
where $\gamma$ is the mean angle between the angular momentum being mixed outward and the
local misaligned angular momentum, and $\beta$ is the local misalignment angle.   Testing this model
against their simulation data, they found excellent agreement.   The expression for the front speed
can also be rewritten in an entirely equivalent form, in which the integration variable is transformed
to the dimensionless $x = r^\prime/r_f$:
\begin{equation}\label{eqn:frontspeed}
\frac{d r_f}{dt} = \langle \cos\gamma\rangle r_f \Omega_{\rm prec}(r_f) \int_0^1 \, dx \, x^{-3/2}
           \frac{\sin\beta(x)}{\sin\beta(r)} \frac{\Sigma(x)}{\Sigma(r)}.
\end{equation}
In other words, the characteristic speed of the alignment front is $\sim r_f \Omega_{\rm prec}$.

\cite{SKH13b} suggested that the place where the front should stall is defined by the place at which
the speed of the alignment front is matched by the speed with which misaligned angular momentum from
the outer disk can be mixed inward.   From their simulation data, it was impossible to estimate the speed
of inward mixing.   However, they proposed a simple parameterization for describing it, in which the
speed of misaligned angular momentum flow was taken to be $r_f/t_{\rm in}$.   In this language, the
steady-state orientation transition would be found at $r_f= R_T$, defined by
\begin{equation}\label{eqn:implicitR_T}
\Omega_{\rm prec}(R_T) t_{\rm in} = \left( \langle\cos\gamma\rangle {\cal I}\right)^{-1},
\end{equation}
where ${\cal I}$ is the dimensionless integral in equation~\ref{eqn:frontspeed}. 

Our simulation data give the location of $R_T$, so equation~\ref{eqn:implicitR_T} can be inverted to find
the characteristic inflow time for misaligned angular momentum:
\begin{equation}
t_{\rm in} = \left( \langle\cos\gamma\rangle {\cal I} \Omega_{\rm prec} (R_T) \right)^{-1}.
\end{equation}
Taking $R_T \simeq 7$, we find that during the period $18 \lesssim t \lesssim 25$, when the alignment
front appears to be almost constant in position, $t_{\rm in} \simeq 11.5$~fiducial orbits, the exact
value varying slightly from time to time.   As a standard of
comparison, we note that the orbital period at $r = 7$ is $\simeq 0.6$~fiducial orbits.

One possible interpretation of this mixing time is in terms of a diffusion model \citep{PP83,Pringle92}, although
\cite{SKH13a} showed that this is not a perfect match with actual time-dependent response to disk warping.
However, even if it does not describe time-dependent effects, a diffusion model might be an appropriate device
with which to model the radial angular momentum mixing supporting a time-steady alignment front.
If one does describe the mixing as a diffusion process,  the associated diffusion coefficient should be
$\sim c_s^2/\Omega$ because the characteristic speed of radial flows $v_r \sim c_s$ \citep{SKH13a},
and orbital mechanics limits the distance such a flow can travel radially to $\sim v_r/\Omega \sim h$.
Although disk warp does create radial pressure gradients of the sort envisioned by \cite{PP83}, the
associated radial velocities are {\it not} regulated by an ``isotropic $\alpha$ viscosity" \citep{SKH13b,Fragile14a},
but by hydrodynamic expansion, gravity, and weak shocks \citep{SKH13a}.
The characteristic speed of mixing is then $\simeq \Phi (h/r)^2 (r/\Delta r) r \Omega(r)$, where $\Delta r/r$ is the
characteristic lengthscale for orientation change within the front, and $\Phi$ is a number of order unity that can
be determined from simulations like ours.   Defining $\Delta r \equiv |\Delta \hat \ell|/|d\hat \ell /d r|$ for
$\Delta \hat \ell$ the difference in direction between the disk's initial orientation and its aligned direction, we find
$\Delta r \simeq 6$.   Equating this speed to the unimpeded $dr_f/dt$ leads to an estimate for the stationary front location
\begin{equation}
R_T \simeq R_* \left[ \langle\cos\gamma\rangle {\cal I} \frac{\Omega_{\rm prec}(R_*)}{\Omega(R_*)}
                \frac{\Delta r/r}{\Phi (h/r)^2}\right]^{2/3}.
\end{equation}
In order to find a scale for $R_T$, we have created units relative to a fiducial radial scale $R_*$.
Note that in terms of the formalism created by \cite{NP00}, this expression corresponds approximately to the
geometric mean of $\tau_{\rm SF 1}$ and $\tau_{\rm SF 2}$.

If this model correctly describes the dynamics of alignment front equilibrium, the order unity quantity $\Phi$ is
\begin{equation}
\Phi \simeq \langle \cos\gamma \rangle {\cal I} \left[ \Omega_{\rm prec}(R_T)/\Omega(R_T)\right] (\Delta r/r) (h/r)^{-1}.
\end{equation}
Our data indicate that $\Phi \simeq 0.6$--0.8, depending on the time at which it is measured.

It is possible that $\Phi$ might be independent of other parameters; however, that remains to be demonstrated.
It could be, for example, that $\Phi$ is dependent on $c_s/v_{\rm orb} \simeq H/r$.   It should also be noted that
because the width of the orientation transition was smaller than the disk simulated, the transition thickness
$\Delta r$ and the internal warp rate $\phat$ were free to adjust to whatever the dynamics required.   There may
be cases in which the intrinsic disk obliquity is large enough, and the radius at which the disk is fed is small enough,
that additional constraints are placed on $\Delta r$ and therefore $\phat$; these might then also alter $\Phi$.
In extreme cases, shock heating might be strong enough to alter $c_s$, leading to nonlinearity in the effective
diffusion rate.

\subsection{What is the role of bending waves?}

As discussed in Sec.~\ref{sec:torque_evol}, we see clear evidence for nonlinear bending wave pulses moving
through the aligning disk.   However, these pulses propagate in a rather irregular way, their amplitudes fluctuating
strongly as they move outward through the disk.    In our previous paper \citep{SKH13b}, we showed that
in a purely hydrodynamic situation these pulses propagate much more regularly.  We attributed the contrast
to the fact that warped hydrodynamic disks are laminar, whereas warped MHD disks are turbulent; the
turbulence disrupts, but does entirely destroy, these waves.   One consequence (as seen both in the
previous simulation and in this one) is that an MHD disk is not able to maintain a state of solid-body rotation,
which it does achieve in a purely hydrodynamic state \citep{SKH13b}.

Since the work of \citet{PP83}, it has been customary to categorize the dynamics of warped disks in terms of
a division between a ``bending wave" regime, in which the ratio of vertically-integrated accretion stress to
vertically-integrated pressure $\alpha$ is $< H/r$, and a ``diffusive" regime, in which $\alpha > H/r$.   It might
then seem a natural question to ask how the phenomena we observe may be classified in these terms.
Before doing so, however, it is useful to recall the origin of this distinction.   It was derived by assuming that linear
bending waves damp at a rate $\alpha\Omega$, while the speeds of the radial motions induced by the warp-created radial
pressure gradients are controlled by an ``isotropic viscosity" acting to reduce vertical shear.   ``Isotropic" meant that
if the vertical shear were the same as the orbital shear, the $r$-$z$ component of the stress would be the same as the
accretion stress, the $r$-$\phi$ component.

However, neither of these assumptions fits well with the conditions in warped MHD disks.   In the vicinity of an alignment front,
the warp is almost always nonlinear in the sense that $\hat\psi > 1$ because even the transonic radial motions induced
by nonlinear warps cannot spread the front enough to make it linear.    This condition is generic because the location
of the front is determined by the condition that the mixing timescale is comparable to the precession timescale.  Indeed, the only
regions in our simulation in which the warp is linear are at sufficiently large radius that they are essentially untouched by
the torques.   Consequently, the relevant damping rate is the one applicable to nonlinear waves.
As shown in \citet{SKH13a}, damping of nonlinear warps takes place very rapidly, and the mechanism is exactly
what is commonly associated with the ``diffusive" regime: radial motions
driven by the radial pressure gradients associated with the warp mix differently-aligned angular momentum across
a radial orientation gradient.   Thus, nonlinear bending waves combine aspects of these two, nominally contrasting,
regimes.

In addition, as shown in \cite{SKH13b} and \cite{Fragile14a}, there is no ``isotropic viscosity".   The $r$-$z$ component
of the Maxwell stress does not behave like a viscosity: it is as likely to strengthen shear as to weaken it.
Moreover, its magnitude, when measured in pressure units, is typically three orders of magnitude smaller than the
stress associated with accretion.    Consequently, the speeds of the radial motions are limited not by a ``viscosity", but,
when the warp is nonlinear, by the dynamics of free expansion in the presence of gravity and by weak shocks that
occur when the radial motions encounter other parts of the warped disk.
For this reason, the radial speeds in regions of nonlinear warp are in general transonic, with Mach numbers that
increase with greater $\hat\psi$ \citep{SKH13a}.    Thus, it is questionable whether the ``bending wave" vs.
``diffusive" distinction can be applied here; the only distinction having the potential to be qualitative
is between linear and nonlinear bending waves.

\subsection{Do warped disks break?}

When disks subjected to Lense-Thirring torques are simulated in pure hydrodynamics with an assumed
isotropic viscosity and using the SPH algorithm, under certain circumstances breaks have appeared between
the outer misaligned section and the inner aligned portion \citep{Nixon2012a,Nixon2012b,Nealon2015}.
One proposed criterion for this to happen is that the external torque ${\cal T}$ exceed the inter-ring torque associated
with accretion stresses $G$ \citep{Nixon2012b,Nealon2015}.   Alternatively, it may be that the criterion is for
the precession timescale to be shorter than the radial sound-crossing timescale \citep{Nealon2015}.

\begin{figure}
\begin{center}
\includegraphics[width=\textwidth,angle=90]{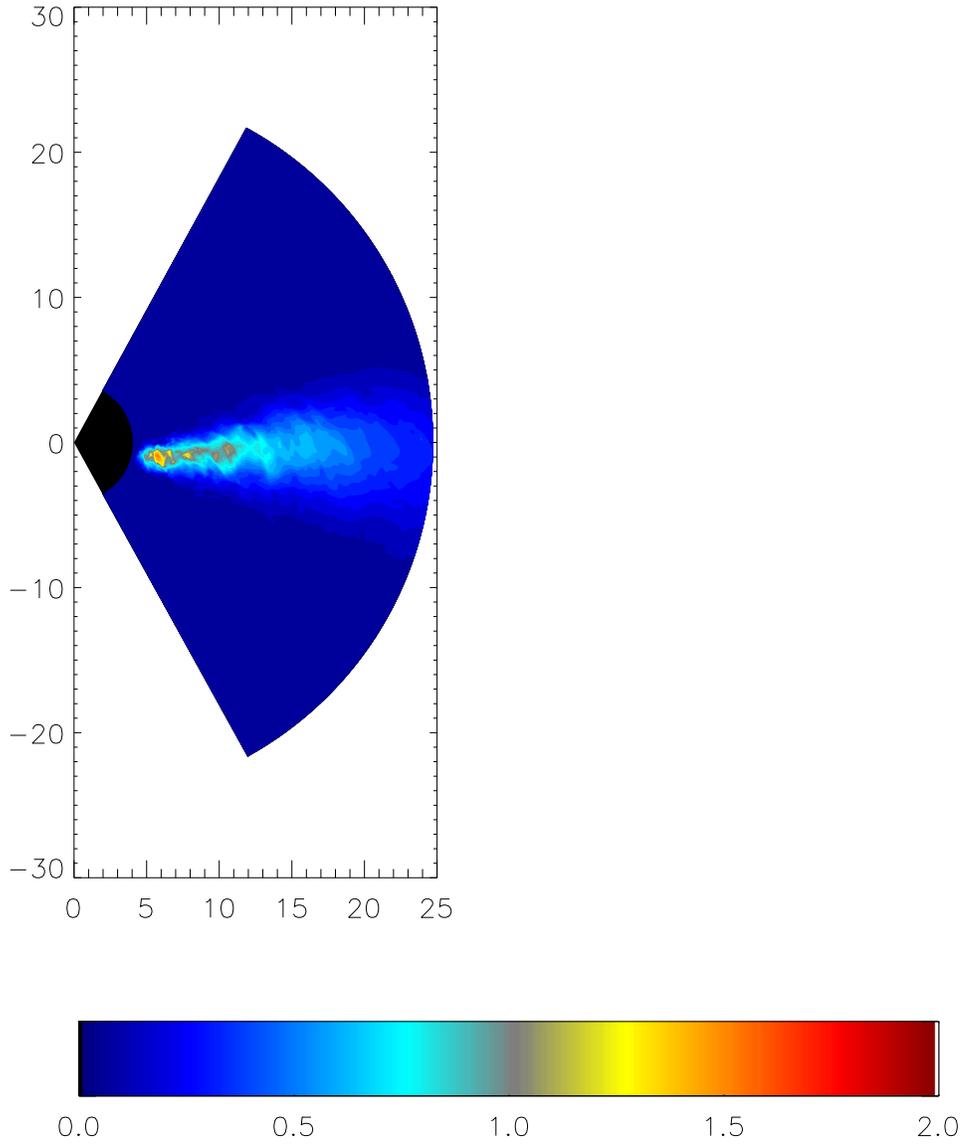} \\
\caption{Linear color contours (see color bar) of the disk density in a poloidal slice at $\phi=0.3\pi$, the approximate
precession angle of the disk at $t=24$.}
\label{fig:poloidal}
\end{center}
\end{figure}

\begin{figure}
\begin{center}
\includegraphics[width=0.6\textwidth,angle=90]{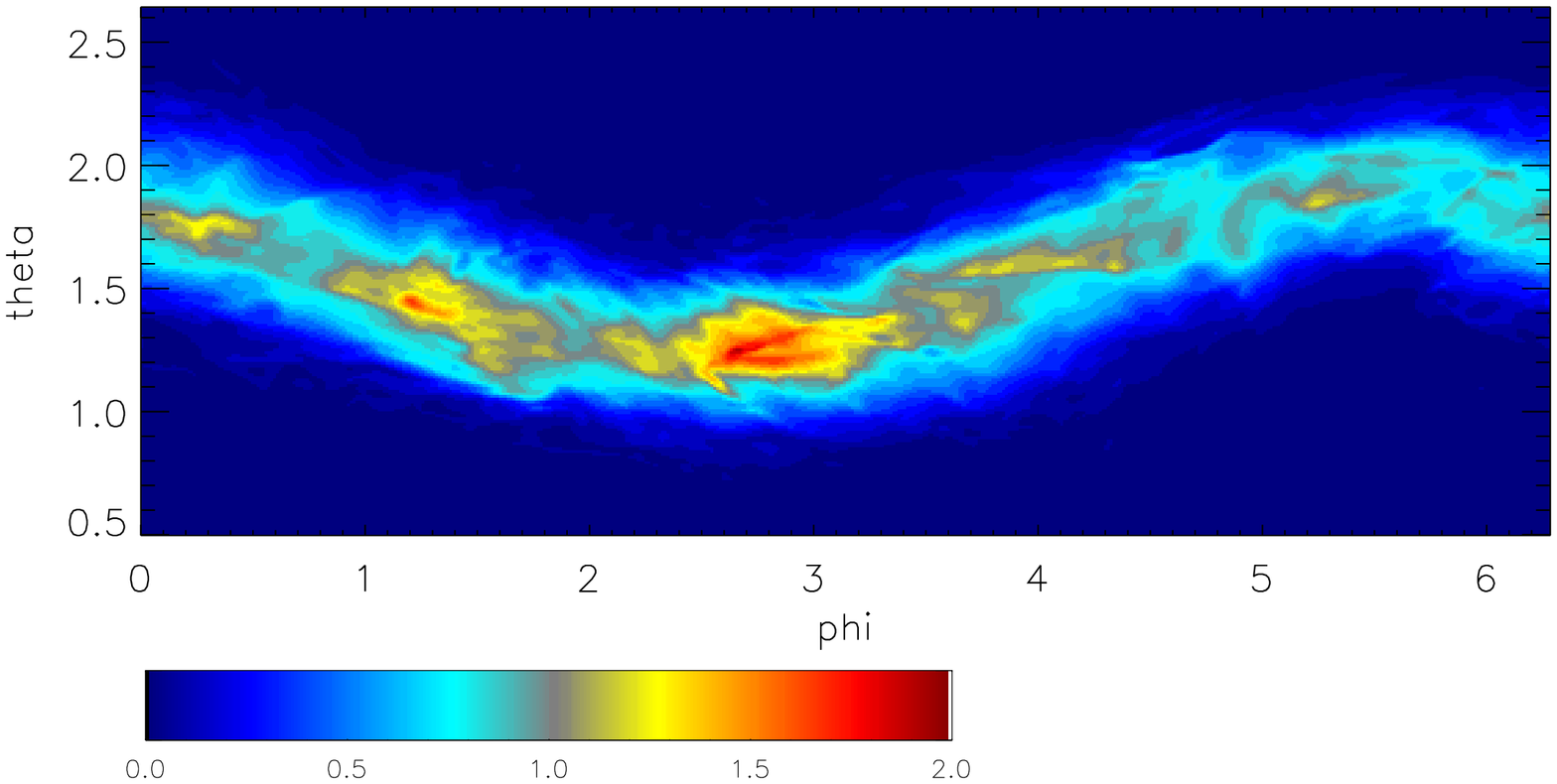} \\
\caption{Linear color contours (see color bar) of the disk density at $t=24$ on the radial shell at $r=7.5$, the radius of sharpest
warp.}
\label{fig:shell}
\end{center}
\end{figure}

We use several diagnostics to test our late-time data for disk breaks in this simulation.   The first is continuity in the
azimuthally-averaged surface density.   As can readily be seen in Figure~\ref{fig:surfdens_late},
$\Sigma(r)$ varies quite smoothly with radius.   Indeed, the surface density near the alignment
transition hardly changes over the $\simeq 15$ fiducial orbits in which the torques operate,
while the surface density at somewhat larger radii increases.     Another is the density distribution
in the poloidal plane measured at the azimuth corresponding to the outer disk precession angle.
This is shown in Figure~\ref{fig:poloidal}.   The plane of the disk does vary, of course.   At small radii,
where it is aligned normal to the black hole spin, it is tilted by $12^\circ$ with respect to the grid; at
large radii, precession around the black hole spin direction moves it out of the equatorial plane.
Throughout the radial extent of the disk, the density distribution is lumpy, but the maximum contrast in density within
the midplane is only about a factor of 2.   Thus, this view also provides no evidence of breaks.   A third diagnostic
is the density distribution on a spherical shell at the radius where the warp is greatest (Fig.~\ref{fig:shell}).
Because the mean angular momentum on this shell is tilted by a bit more than half the initial inclination relative to the coordinate
equatorial plane, and it has precessed, the disk plane traces a sinusoidal pattern in polar angle relative
to the $\theta\times\phi$ plane in this coordinate system.   It shows density fluctuations at the factor of 3 level, but there is
also no indication of a dramatic break here either, even though this is the radius where $\hat\psi$ is
greatest (see Fig.~\ref{fig:psihat}).

It is therefore instructive to test whether the proposed break criteria apply to a calculation in which all
stresses are physical and numerical transport
can be readily controlled.   The first criterion can be concisely phrased if the actual accretion stress is
measured in pressure units, i.e., in terms of $\alpha$, the time-averaged ratio of vertically-integrated
stress to vertically-integrated pressure.  It is then
\begin{equation}
\frac{\cal T}{G} = \frac{(\Omega_{\rm prec}/\Omega)\sin\theta}{\alpha (h/r)^2}.
\end{equation}
At the location of the orientation transition ($r \simeq 7$), the frequency ratio is $\simeq 0.1$, while
our measured stress ratio $\alpha$ declines with radius from $\simeq 0.1$ at $r \simeq 5$ to $\simeq 0.05$
at $r \simeq 10$.    Combined with our disk aspect ratio $h/r \simeq 0.1$
and the outer disk inclination $\sin\theta \simeq 0.2$, we find ${\cal T}/G \simeq 27$.    Thus, even though
the precessional torque is considerably greater than the inter-ring accretion torque, the disk does not break.

One possible explanation for this discrepancy is that the argument for the ${\cal T}/G$ criterion
assumes that ordinary accretion stresses are responsible for restoring matter to regions in which
the precessional stress might otherwise create gaps.   However, warped disks induce strong
Reynolds stresses, and these can easily be much stronger than the Maxwell stresses due to
correlated MHD turbulence that ordinarily drive accretion.   In fact, at a steady-state alignment
transition front, these Reynolds stresses produce a torque that is---by definition---equal to the
aligning torque at the front, and the aligning torque is comparable to the full Lense-Thirring
precessional torque.    Thus, if the correct criterion for producing a break is for
the Lense-Thirring torques to exceed the smoothing torque due to internal stresses, there could
never be a strong break at a disk's most vulnerable point, the alignment transition.

The second criterion amounts to computing the product
\begin{equation}
\Omega_{\rm prec} t_{\rm cross}  = \frac{\Omega_{\rm prec}/\Omega}{h/r},
\end{equation}
where $t_{\rm cross}$ is the time for sound waves to traverse a distance comparable to
the radius where this product is measured.    If the theory for the location of the transition radius
outlined in the previous section is correct, this criterion may be rewritten as
\begin{equation}
\Omega_{\rm prec} t_{\rm cross}  = {\Phi \over \langle\cos\gamma\rangle {\cal I} (\Delta r/r)}.
\end{equation}
Thus, this ratio is always $\sim {\cal I}^{-1}$.   For our simulation's parameters, it is $\simeq 1$
at the alignment transition.    Just as for the previous criterion, our theory predicts that this
criterion, too, should always be marginal, and therefore strong breaks would not be expected.

\cite{Nealon2015} suggest that this criterion applies in what they deem the ``bending wave" regime, a regime in
which the pressure-coupling of bending waves provides the strongest coupling between adjacent rings.   Because
these waves generically travel at $0.5c_s$, the sound wave crossing time is a reasonable estimate of their
crossing time also.   As we have already argued, linear bending waves do not appear to be particularly
important contributors to tilted disk dynamics.   However, nonlinear warps result in transonic flows, whose
effects also travel at $\sim c_s$.   In this sense, it is not surprising that we would predict that this criterion,
like our reworked version of the first, should also always be marginal at the orientation transition radius.

\subsection{Is apsidal precession relevant?}

It has been argued in a number of places \citep{II97,Fragile14a,Nealon2015} that apsidal precession can
lead to radial alignment oscillations, perhaps affecting overall disk alignment.   Nonetheless, in our lowest-order
PN accounting of general relativistic effects, we have included the gravitomagnetic term, but not the term associated
with apsidal precession, even though the two effects are formally the same order in the expansion.
We did so for a number of reasons.   One reason is that there is no way to make the apsidal precession
consistent with the magnitude of our amplified gravitomagnetic term.   However, we also believe it to have at
most minor physical influence.   Several arguments point in this direction.   First, the amplitude of the velocity
perturbations created by this effect is very small.    They are proportional to the eccentricity $e$ of the
orbits, and $e \sim c_s/v_{\rm orb}$.   As a result, the Mach number of these motions is only of order the
ratio of the gravitational radius $r_g$ to the radius, which is generally quite small for $r \sim R_T$.   They are then
far slower than the transonic radial motions themselves; in fact, they are very small even compared
to the fluid motions associated with the turbulence, which are $\sim 0.1 c_s$.   Second, the radial flows
are altered by hydrodynamics on the orbital period, as streams encounter other portions of the disk, sometimes
through shocks.   By contrast, during one orbital period, the magnitude of the apsidal
precession angle is only $6\pi r_g/r \ll 1$ for $e \ll 1$.    Thus, during the time over which a specific elliptical
orbit exists, it is rotated by only a very small amount, while that fluid's motions are changed at the order-unity
level by other mechanisms.   Third, the apsidal precession can also be viewed as
due to a small offset between the frequency of radial epicyclic motions and the orbital frequency.   However,
in an MHD orbiting fluid, the magnetorotational instability makes radial epicyclic motions unstable, and the
fastest growing mode has a growth rate $\sim \Omega$.    It is hard to see how the apsidal precession
would be significant relative to all these other effects.

\section{Conclusions}\label{sec:con}


Our simulation has shown that an accretion disk whose intrinsic orbital plane is inclined obliquely to the equatorial
plane of the central spinning black hole can achieve a steady-state in which the inner disk aligns with the black hole's
equatorial plane while the outer disk retains its original orientation.

Using the alignment front propagation formalism developed in a previous paper \citep{SKH13b},
we have, for the first time, measured directly the timescale for misaligned angular momentum mixing due to the
radial flows induced by disk warping.   In this specific case, that time scale is $\simeq 12$ local orbital periods.
In the context of a time-steady alignment front, it is possible to estimate this mixing timescale in terms of a
diffusion model of the sort first explored by \citet{PP83} and \citet{Pringle92}.  Because nonlinear warps
generically produce transonic radial motions, one would expect the diffusion coefficient in such a model to be
$\Phi c_s^2/\Omega$, where $\Phi$ is a dimensionless number of order unity.    Interpreting the mixing time
we measure in terms of such a model, one would then expect $t_{\rm in}^{-1} = \Phi (h/r)^2 (\Delta r/r)^{-1} \Omega$
for fractional front width $\Delta r/r$.    Calibrating this relation with our simulation data, we find that
$\Phi \simeq 0.6$--0.8 for our measured fractional front width $\Delta r/r \simeq 6/7$.  Finding a coefficient of
order unity is consistent with the thought that a diffusion model of this character may be appropriate to estimating
the equilibrium properties of disk warps, even though this model is inadequate to describe their time-dependent
properties.    Stronger evidence for this proposition
must await simulations with different values of the sound speed to orbital speed ratio and perhaps also
different disk inclination angles; it remains possible that $\Phi$ could depend on these, or other, parameters.

Nonetheless, if this estimate is good to factors of several, it permits estimation
of steady-state orientation transition radii, given the black hole mass and spin and the accretion disk surface
density and aspect ratio profiles.   In fact, this approach has already been taken by \citet{MK2013}, who
assumed $\Phi \simeq 1$; the results of our new simulation support their assumption.

We have also probed several quantitative aspects of the alignment front's approach to equilibrium.
For example, we observed an overshoot in the front's location before settling into a steady-state.   A pure
diffusion picture of warp motion would not yield this sort of behavior, but its qualitative character (in particular,
its relaxation time) is consistent with previous detailed studies of warped disk hydrodynamics \citep{SKH13a}.
We have also begun to constrain suggestions that disks undergoing Lense-Thirring torques can suffer sharp
breaks \citep{Nixon2012a,Nixon2012b}.  \citet{Nealon2015} offered criteria for when these breaks should
develop; despite satisfying one of their criteria strongly and another one marginally, we find
no evidence for this happening.    If their primary criterion were reframed to include the effects of the
radial motions induced by a strong warp, we would find both are marginal, and for the same reason:
at a steady-state alignment front, the torque due to
the Reynolds stress of these motions is automatically matched to the torque applied by gravitomagnetism,
thereby creating a relatively rapid change in disk plane orientation across the front, but not a sharp break.

A number of questions remain open for future work.   Among the most interesting are the degree to which
$\Phi$ may vary with parameters such as $c_s/v_{\rm orb}$ and the possible effect of boundary conditions
that prevent the transition front from extending across as wide a radial range as internal dynamics would
demand.   If the location of disk-feeding is at a relatively small radius and the total change in angle across
the alignment front is relatively large, the warp rate $\phat$ may be forced to become larger than it would
be otherwise.   In such a situation, there might be both a significant alteration in $\Phi$ and enough shock
heating to create nonlinearity through thermodynamic effects.

\section*{Acknowledgements}
We thank John Papaloizou for enlightening and encouraging conversations.
This work was partially supported under National Science Foundation grants AST-1028111 (JHK) and
AST-0908869 (JFH), and NASA grant NNX14AB43G (JHK and JFH).
The National Science Foundation also supported this research in part through XSEDE resources
on the Stampede cluster through allocation TG-MCA95C003.

\bibliography{Bib}

\end{document}